\def\be{\begin{equation}}
\def\ee{\end{equation}}
\def\ba{\begin{array}}
\def\ea{\end{array}}

\documentclass{article}
\usepackage{titling}
\usepackage{multicol}
\usepackage{amsfonts}
\usepackage{amssymb}
\usepackage{amsmath,amssymb,amstext,amsfonts,array,hhline,tabularx,graphics}
\usepackage{latexsym}
\usepackage{arydshln}
\usepackage{epsfig,graphicx, color}
\usepackage{amsmath}
\usepackage{abstract}
\usepackage{cite}

\newtheorem{theorem}{Theorem}
\newtheorem{lemma}{Lemma}

\newtheorem{corollary}{Corollary}

\begin{document}
\parskip=3pt
\parindent=18pt
\baselineskip=20pt \setcounter{page}{1}

 \title{\large\bf Uncertainty Relations based on Wigner-Yanase Skew Information}
\date{}
\author{Xiaofen Huang$^{1}$, Tinggui Zhang$^{1\ast}$, Naihuan Jing$^{2,3}$\\[10pt]
\footnotesize
\small 1 School of Mathematics and Statistics, Hainan Normal University, Haikou, 571158, China\\
\small 2 Department of Mathematics, Shanghai University, Shanghai 200444, China\\
\small 3 Department of Mathematics, North Carolina State University,
Raleigh, NC 27695, USA\\}

\date{}

\maketitle

\centerline{$^\ast$ Correspondence to tinggui333@163.com}
\bigskip

\begin{abstract}
In this paper, we use certain norm inequalities to obtain new
uncertain relations based on the Wigner-Yanase skew information. First for arbitrary finite number of observables we derive an
uncertainty relation outperforming previous lower bounds. We then propose new weighted uncertainty relations
for two noncompatible observables. Two
separable criteria via skew information are also obtained.
\end{abstract}


%\keywords{Uncertainty relation; Entanglement;}

\begin{multicols}{2}

\section{Introduction}
\noindent
Uncertainty relation is one of the fundamental building
blocks of quantum theory, and plays an significant role in quantum
information and quantum mechanics %\cite{fou1-fou4}
\cite{fou1, fou2, fou3, fou4}. It reveals a fundamental limit
with which certain pairs of physical properties of a
particle such as position and momentum cannot be simultaneously
known exactly.

The uncertainty relations dominated the developments of physics that ranges from foundations to quantum information, quantum communication and other areas as well, which gives rise
to wide applications in entanglement detection \cite{entan1, entan2}, as well as security analysis of quantum key distribution in quantum cryptography \cite{secu}, quantum metrology
and quantum speed limit \cite{qsl1, qsl2, qsl3}.

Generally the uncertainty relations are expressed in terms of the product of variances of the measurement results of two incompatible observables\cite{sk2}. Besides variance based uncertainty,
there are also other ways to formulate the principle, such as in terms of entropies \cite{ent1, ent2, ent3, sk3, sk4}, majorization \cite{maj1, maj2, maj3, qiao} and there are also fine-grained
uncertainty relations \cite{fin1, fin2, yu}.

The quantum uncertainty relation can be also described in terms of
skew information\cite{sk1}. In this work we will only focus on the skew
information-based additive uncertainty relations. In 1963, Wigener and
Yanase \cite{WY} introduced the skew information $I_{\rho}(H)$ of the observable $H$
as a  measure of quantum information contained in a state $\rho$,
namely,
\begin{equation} \label{def}
  \begin{split}
I_{\rho}(H)&=-\frac{1}{2}\rm {Tr}([\sqrt{\rho}, H]^2)\\
&=\rm {Tr}(\rho H^2)-\rm {Tr}(\sqrt{\rho}H\sqrt{\rho}H).
\end{split}
\end{equation}
In addition, the skew information can be cast as the norm form
according to the Frobenius norm $\parallel \ \ \parallel$, that is
\begin{equation}
I_{\rho}(H)=\frac{1}{2}\parallel[\sqrt{\rho}, H]\parallel^2.
\end{equation}
In this formulism $I_{\rho}(H)$ can be viewed as a kind of degree for
non-commutativity between the quantum state $\rho$ and the observable
$H$. It manifestly vanishes when $\rho$ commutes with $H$, and it is
homogeneous in $\rho$. By means of the skew information and the
decomposition of the variance, a stronger uncertainty relation was
presented for mixed states \cite{luos,chen}. Since information is
lost when separated systems are united such a measure should be
decreasing under the mixing of state \cite{Lieb}, that is, convex
in $\rho$.

The Wigner-Yanase skew information has becomes a useful tool in quantum
information theory, for instance, characterizing entanglement \cite{czq},
begging a measure of the $H$ coherence of the state $\rho$, and
quantifying the dynamics of some physical phenomena.
In this paper, we present more tighter uncertainty relations based on the Wigner-Yanase skew
information, and the newly given uncertainty principle is shown to be applicable to judge separability.

\section{Uncertainty Relation Based on Skew Information for Multi Operators}
\noindent
In this section, we present an uncertainty relation based
on the skew information for multiple incompatible observables.

\begin{theorem} For noncommutative observables $A_i$, $i=1, 2, ..., n$, the following uncertainty inequalities hold
\begin{equation}\label{rest2}
\begin{split}
 \sum_{i}^{n}I_{\rho}( A_i)&\geq\frac{1}{n} I_{\rho}(\sum_{i}^{n} A_i)\\
&+\frac{1}{n^2}\bigg(\sum_{1\leq i<j\leq n}\sqrt{I_{\rho}( A_i- A_j)}\bigg)^2.
  \end{split}
  \end{equation}
If $A_{i}$s are mutually noncommutative, then the lower bound in (\ref{rest2}) is nonzero.
\end{theorem}
\textbf{Proof:} On a Hilbert space the following identity holds
\cite{bonr}:
\begin{equation}
n\sum_{i=1}^n\parallel u_i\parallel^2=\parallel \sum_{i=1}^n u _i\parallel^2+\sum_{1\leq i<j\leq n}\parallel u_i-u_j\parallel^2,
\end{equation}
where $u_i$ is a vector in Hilbert space.

Also the inequality holds
 $$
 \sum_{1\leq i<j\leq n}\parallel u_i-u_j\parallel^2
 \geq \frac{1}{n}(\sum_{1\leq i<j\leq n}\parallel(u_i-u_j)\parallel)^2.
 $$
Therefore one has that
\begin{equation}
  \begin{split}
 \sum_{i=1}^n\parallel u_i\parallel^2&\geq \frac{1}{n}\parallel \sum_{i=1}^n u _i\parallel^2 \\
 &+\frac{1}{n^2}(\sum_{1\leq i<j\leq n}\parallel(u_i-u_j)\parallel)^2.
  \end{split}
  \end{equation}
Let $u_i=[\sqrt{\rho}, A_i]$, we obtain the uncertainty relation for
skew information in the form (\ref{rest2}). Moreover, both
$I_{\rho}(\sum_{i}^n A_i)$ and $I_{\rho}(A_i-A_j)$ are equal to
zero if the lower bound $(\ref{rest2})$ is zero, which implies that
$I_{\rho}(A_i)=0$, then the observes $A_i$ are mutually
commutative.

As the operators $A_{i}$'s are mutually noncommutative,
the inequality (\ref{rest2}) is nontrivial, so the lower bound (\ref{rest2}) is nontrivial.

\textbf{Remark 1:} In the case of pure state $\rho$ which is an eigenvector of observable $A$, the skew information
 $I_{\rho}(A)=0$. This means that the sum of skew information $I_{\rho}(A)+I_{\rho}(B)$ is nontrivial if $\rho$ is not a common eigenvector of observables $A$ and $B$. However,
 both Heisenberg-Robertson's and Schr\"{o}dinger's uncertainty relations are trivial in that case.

\textbf{Remark 2:} In particular, if $\rho$ is a pure state, the skew information $I_{\rho}(H)$ happens to be the variance $(\triangle_{\rho} H)^2$. According to the definition
of skew information $I_{\rho}(H)=-\frac{1}{2}\rm {Tr}([\sqrt{\rho}, H]^2)=\rm {Tr}(\rho H^2)-\rm {Tr}(\sqrt{\rho}H\sqrt{\rho}H)$, in case of $\rho=|\varphi\rangle \langle \varphi |$,
then $I_{\rho}(H)=\langle\varphi |H^2|\varphi\rangle-\langle \varphi|H|\varphi \rangle^2=(\bigtriangleup_{\rho} H)^2$. Thus, our relation happens to be the inequality obtained by Song in \cite{song},
\begin{equation} \label{song}
  \begin{split}
  \sum_{i}^{n}\bigtriangleup_{\rho}( A_i)^2 & \geq\frac{1}{n}\bigg( \bigtriangleup_{\rho}(\sum_{i}^{n} A_i)\bigg)^2 \\
  &+ \frac{1}{n^2}\bigg(\sum_{1\leq i<j\leq n}\bigtriangleup_{\rho}( A_i- A_j)\bigg)^2.
  \end{split}
\end{equation}
It means that the relation (\ref{rest2}) can reduce to the inequality (\ref{song}) in case of pure states.

When there are two noncommutative observables in Theorem 1, we can
get a corollary below.
\begin{corollary} For noncommutative observables $A$ and $B$, we have
\begin{equation}\label{co}
\begin{split}
  I_{\rho}(A)+I_{\rho}(B) & \geq\frac{1}{2}I_{\rho}(A+B)+\frac{1}{4}I_{\rho}(A-B) \\
  & \geq \frac{1}{2} I_{\rho}(A+B).
\end{split}
\end{equation}
\end{corollary}

 In case $\rho$ is a pure state, we can rewrite the inequality (\ref{co}) according to the relation $I_{\rho}(H)=(\bigtriangleup_{\rho} H)^2$, thus we obtain an inequality based on variance in the following
\begin{equation}\label{co1}
\begin{split}
(\bigtriangleup_{\rho} A)^2+(\bigtriangleup_{\rho} B)^2&\geq \frac{1}{2}\bigg(\bigtriangleup_{\rho}(A+B)\bigg)^2\\
&+\frac{1}{4}\bigg(\bigtriangleup_{\rho}(A-B)\bigg)^2.
\end{split}
\end{equation}
Also our relation (\ref{co1}) has a stronger lower bound,  which is tighter than the uncertainty relation derived by Maccone1 and Pati in \cite{pa}:
%\begin{equation}
%(\bigtriangleup_{\rho} A)^2+(\bigtriangleup_{\rho} B)^2\geq \frac{1}{2}\langle\psi ^{\bot}_{A+B}|A+B|\psi\rangle^2=\frac{1}{2} \bigg(\bigtriangleup_{\rho}( A+B)\bigg)^2.
%\end{equation}
\begin{equation}
\begin{split}
  (\bigtriangleup_{\rho} A)^2+(\bigtriangleup_{\rho} B)^2&\geq \frac{1}{2}\langle\psi ^{\bot}_{A+B}|A+B|\psi\rangle^2 \\
  & =\frac{1}{2} \bigg(\bigtriangleup_{\rho}( A+B)\bigg)^2.
\end{split}
\end{equation}

It is worth noting that Chen et al. derived an uncertainty relation based on Wigner-Yanase skew information \cite{chen} which states that
\begin{equation}\label{ch}
\begin{split}
 \sum_{i}^{n}I_{\rho}(A_i)&\geq \frac{1}{n-2}[\sum_{1\leq i<j\leq n} I_{\rho}(A_i+A_j)
 - \\
  & \frac{1}{(n-1)^2}\bigg(\sum_{1\leq i<j\leq
 n}\sqrt{I_{\rho}(A_i+A_j)}\bigg)^2].
\end{split}
\end{equation}

Also, inequality (\ref{rest2}) has a stronger lower bound than the one in (\ref{ch}) for a qubit system\cite{song}. As an example, we consider the Pauli matrices
\begin{equation*}
\begin{split}
&\sigma_1=\left(
  \begin{array}{cc}
    0& 1\\
    1& 0\\
  \end{array}
\right), \sigma_2=\left(
  \begin{array}{cc}
    0& -\rm{i}\\
    \rm {i}&  0\\
  \end{array}
\right), \\
& \sigma_3=\left(
  \begin{array}{cc}
   1 &0\\
   0&-1\\
  \end{array}
\right).
\end{split}
\end{equation*}

Let $\rho=\frac{1}{2}(I+\vec{r}\vec{\sigma})$, where the Bloch
vector $\vec{r}=(\frac{\sqrt{3}}{2}\rm {\cos}\theta,
\frac{\sqrt{3}}{2}\rm {\sin}\theta, 0 )$. Then
$I_{\rho}(\sigma_1-\sigma_2)=\frac{1}{2}(1+\frac{1}{2}\sin2\theta)$,
$I_{\rho}(\sigma_1-\sigma_3)=\frac{1}{4}(3-\cos2\theta)$,
$I_{\rho}(\sigma_2-\sigma_3)=\frac{1}{4}(3+\cos2\theta)$,
$I_{\rho}(\sigma_1+\sigma_2+\sigma_3)=1-\frac{1}{2}\sin2\theta$. The
comparison between the two bounds \eqref{rest2} and \eqref{ch} is
given in Figure 1, where one sees clearly that our bound outperforms
that of (\ref{ch}).

\begin{figure*}[htbp]
\begin{center}
\includegraphics[scale=0.5]{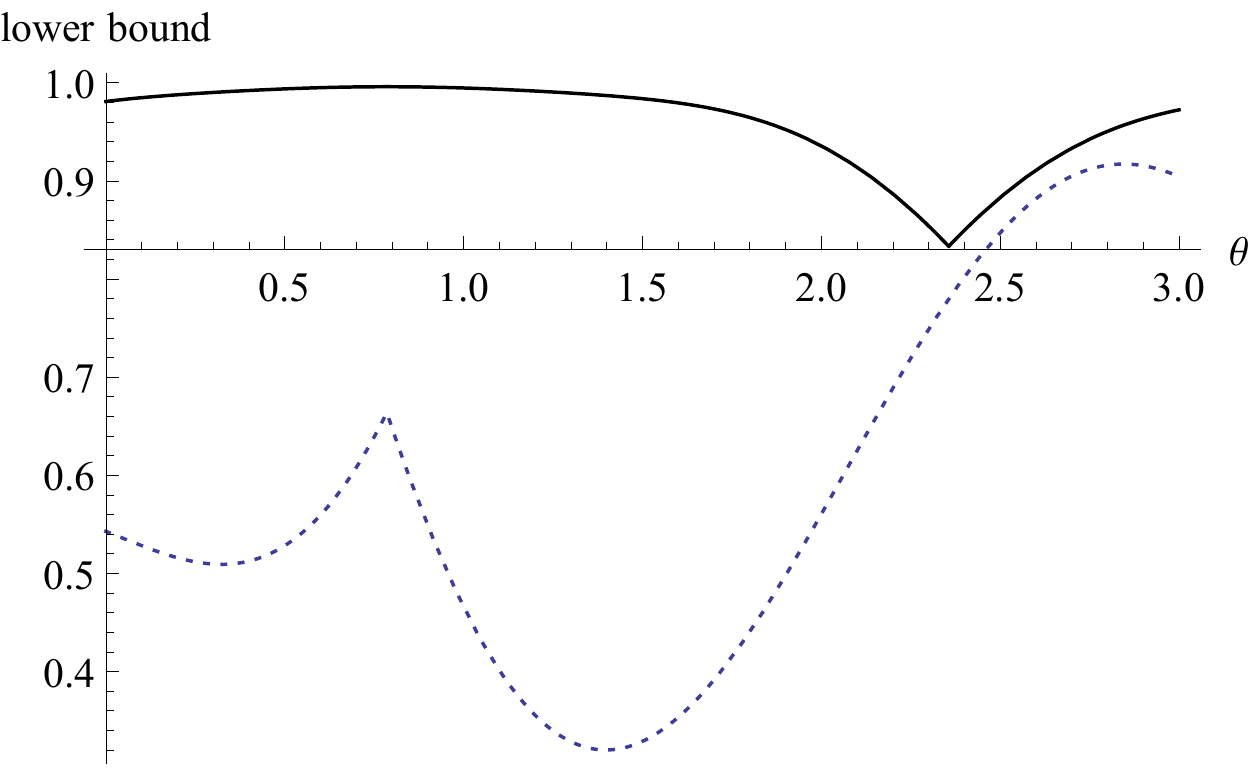}
\end{center}
\caption{Comparison of our bound with that of Chen et al.: The
solid line for the lower bound in \eqref{rest2}, the dotted line
for the lower bound in \eqref{ch}. \label{Fig:1}}
\end{figure*}

Furthermore, our relation (\ref{rest2}) is stronger than the one such as derived from the uncertainty inequality for two
observables\cite{pa}. As an example, we consider another two uncertainty relations derived from the parallelogram law in the Hilbert space: $2(||u||^2+||v|||^2)=|u+v||^2+||u-v||^2$.
Let $u=[\sqrt{\rho}, A]$, $v=[\sqrt{\rho}, B]$, $A$ and $B$ are two incompatible observables, then we get uncertainty relations based on Wigner-Yanase skew information
\begin{equation} \label{pa}
I_{\rho}(A)+I_{\rho}(B)=\frac{1}{2}\Big(I_{\rho}(A+B)+I_{\rho}(A-B)\Big).
\end{equation}

Using the above uncertainty equality, one can obtain two inequalities for arbitrary $n$ observables, namely,
\begin{equation}\label{p1}
\sum_{i=1}^{n}I_{\rho}(A_i)\geq\frac{1}{2(n-1)}\sum_{1\leq i<j\leq n} I_{\rho}(A_i+A_j),
\end{equation}
\begin{equation}\label{p2}
\sum_{i=1}^{n}I_{\rho}(A_i)\geq\frac{1}{2(n-1)}\sum_{1\leq i<j\leq n}I_{\rho}(A_i-A_j).
\end{equation}
%It is noting that the lower bound of (\ref{p2}) is stronger than the one of (\ref{p1}).

There is an example of comparison between our relation (\ref{rest2}) and ones (\ref{ch}),(\ref{p1}). We consider the spin-1 system with the pure state
 $|\psi\rangle=\cos \frac{\theta}{2}|0\rangle+\sin \frac{\theta}{2}|2\rangle$, $0\leq\theta<2\pi$. Take the angular momentum operators\cite{ang} with $\hbar=1$:
\begin{equation}
\begin{split}
&J_{x}=\frac{1}{\sqrt{2}}\left(
                          \begin{array}{ccc}
                            0 & 1 & 0 \\
                            1 & 0 & 1 \\
                            0 & 1 & 0 \\
                          \end{array}
                        \right),
J_{y}=\frac{1}{\sqrt{2}}\left(
                          \begin{array}{ccc}
                            0 & -\rm {i} & 0 \\
                            \rm {i} & 0 & -\rm {i} \\
                            0 & \rm {i} & 0 \\
                          \end{array}
                        \right), \\
& J_{z}=\frac{1}{\sqrt{2}}\left(
                          \begin{array}{ccc}
                            1 & 0 & 0 \\
                            0 & 0 & 0 \\
                            0 & 0 & -1 \\
                          \end{array}
                        \right).
\end{split}
\end{equation}

Direct calculation gives
\begin{equation*}
\begin{split}
&I_{\rho}(J_{x})=\frac{1}{2}(1+\sin \theta),~
I_{\rho}(J_{y})=\frac{1}{2}(1-\sin \theta),\\
&I_{\rho}(J_{z})=\sin^2 \theta,~~ I_{\rho}(J_{x}+J_{y})=1,\\
&I_{\rho}(J_{x}+J_{z})=\frac{1}{2}(1+\sin \theta)+\sin^2 \theta,~\\
&I_{\rho}(J_{y}+J_{z})=\frac{1}{2}(1-\sin\theta)+\sin^2 \theta,\\
&I_{\rho}(J_{x}+J_{y}+J_{z})=1+\sin^2 \theta,~\\
&I_{\rho}(J_{x}-J_{z})= I_{\rho}(J_{y}-J_{z})=\sin^2 \theta,\\
&I_{\rho}(J_{x}-J_{y})=0.~
\end{split}
\end{equation*}

The comparison between the lower bounds (\ref{rest2}), (\ref{ch}) and (\ref{p1}) is shown by Figure 2. The results suggest that the relation (\ref{rest2})
can give tighter bound than other ones ((\ref{ch}) and (\ref{p1}) for a spin-1 particle and measurement of angular momentum operators $J_{x}$, $J_{y}$ and $J_{z}$.

\begin{figure*}[htbp]
\begin{center}
\includegraphics[scale=0.5]{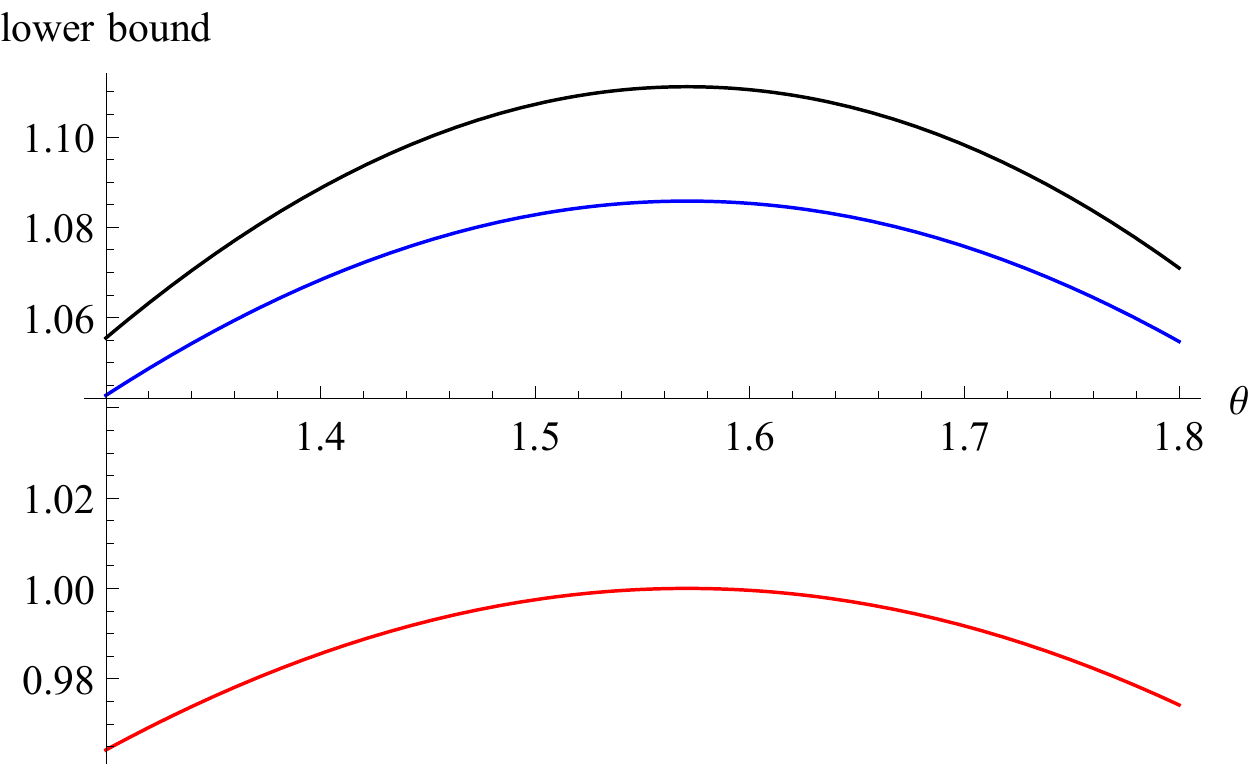}
\end{center}
\caption{Comparison of our bound  (\ref{rest2}) with that of (\ref{ch}) and (\ref{p1}): the
black line for the lower bound in \eqref{rest2}, the blue line
for the lower bound in \eqref{ch}, and the red line is the lower bound for (\ref{p1}). \label{Fig:2}}
\end{figure*}

\section{Uncertainty Relation in term of Skew Information with Weight}
\noindent
Additionally we can get many uncertainty relations if involving parameters, so we consider uncertainty relations with weight based on skew information for two noncommutative observables.

\begin{theorem}For two non-commutativity observables $A$ and $B$, we have the uncertainty relation with weight
\begin{equation}\label{6}
\begin{split}
I_{\rho}(A-B)&+I_{\rho}(\frac{\lambda-1}{\lambda}A-B)\leq
\frac{1}{\lambda}I_{\rho}(A)\\
&+\frac{1}{1-\lambda} I_{\rho}(B)\leq
I_{\rho}(A-B)\\
&+I_{\rho}(A-\frac{\lambda}{\lambda-1}B),
\end{split}
\end{equation}
where $\frac{1}{2}\leq\lambda<1$, and the equality holds when $\lambda=\frac{1}{2}$.
\end{theorem}
\textbf{Proof:} Recall that for bounded linear operators $U$ and $V$ in Hilbert space the following inequalities hold \cite{Bohr}
\begin{equation}
\begin{split}
\|U-V\|^2&+\|(1-p)U-V)\|^2 \leq p\| U\|^2+q \|V\|^2\\
&\leq \|U-V\|^2+\|U-(1-q)V\|^2,
\end{split}
\end{equation}
%\begin{equation}
%\|U-V\|^2+\|(1-p)U-V)\|^2 \leq p\| U\|^2+q \|V\|^2\leq \|U-V\|^2+\|U-(1-q)V\|^2,
%\end{equation}
 for any $1<p\leq 2$, $\frac{1}{p}+\frac{1}{q}=1$, and the equalities hold if and only if $p=2$ or $V=(1-p)U$. Now set $p= \frac{1}{\lambda}$, $U=[\sqrt{\rho}, A]$ and $V=[\sqrt{\rho}, B]$,
 we obtain the uncertainty relation $(\ref{6})$.

\textbf{Remark 3: }In particular, in case of $\lambda=\frac{1}{2}$, inequality (\ref{6}) happens to the parallelogram law in term of skew information (\ref{pa}).

The idea of weighted averaging is one of the popular techniques
in both statistical mechanics and mathematical physics. Through
the weighted averaging one may know better about the whole picture
in an unbiased way. Also we consider a perturbation of $A$ and $B$,
or $A'=\sqrt{1/\lambda}A$, $B'=\sqrt{1/(1-\lambda)}B$, then
$$
I_{\rho}(A')+ I_{\rho}(B')=\frac{1}{\lambda} I_{\rho}(A)+\frac{1}{1-\lambda} I_{\rho}(B).
$$
This means that the lower bound of the sum of skew information can be obtained by scaled observables.

Our lower bound remains nonzero unless $\rho$ is a common eigenvector of $A$ and $B$, which means that besides having a nontrivial bound in almost all cases, our weighted uncertainty relations
 can also lead to a tighter bound for the sum of skew information.

Let us consider again the Pauli matrices $\sigma_1, \sigma_2,
\sigma_3$ and the measured state given by the family of states with
the Bloch vector $\vec{r}=(\frac{\sqrt{3}}{2}\cos\theta,
\frac{\sqrt{3}}{2}\sin\theta, 0 )$, $\theta\in (0, \pi)$. It is
shown in Figure 2 that the sum of skew information with weight
$\lambda I_{\rho}(\sigma_1)+\frac{1}{1-\lambda} I_{\rho}(\sigma_2)$.

\begin{figure*}[htbp]
\begin{center}
\includegraphics[scale=0.5]{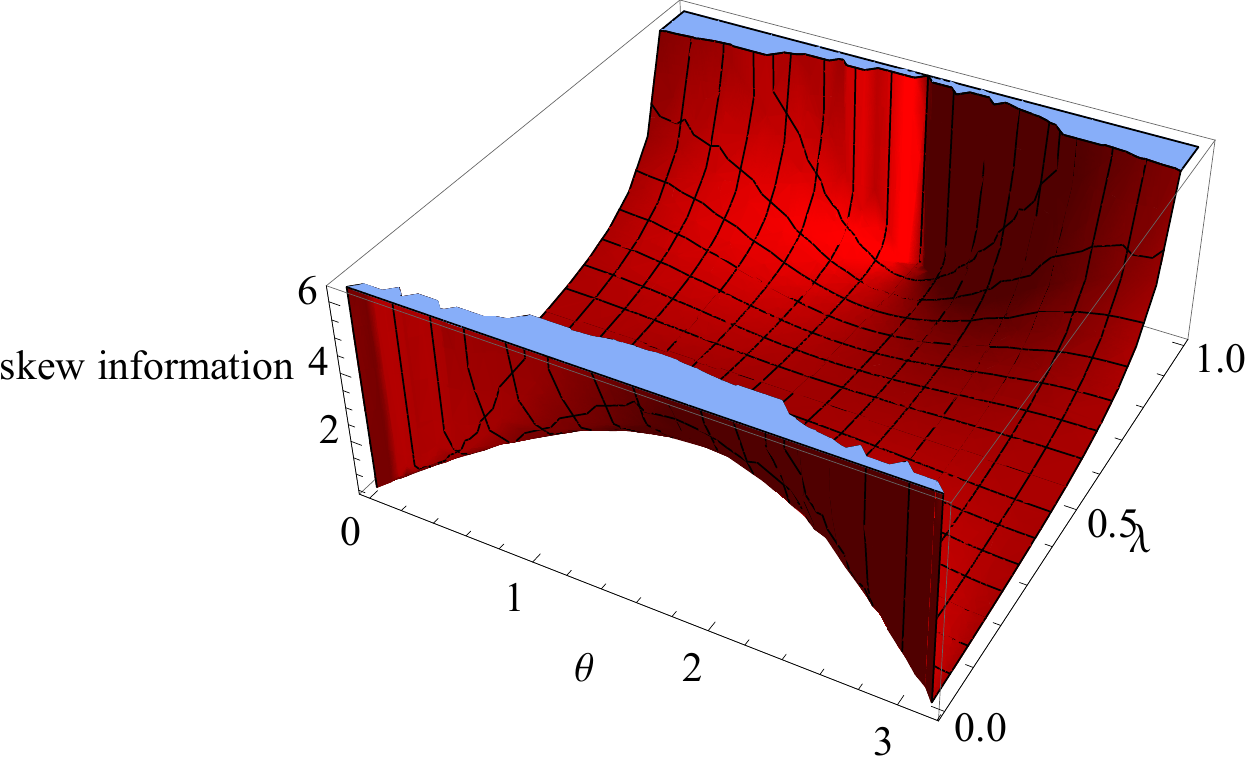}
\end{center}
\caption{The sum uncertainty relations based on skew information with weight are satisfied by observables $\sigma_{1}$ and $\sigma_{2}$ with state $\rho$.}\label{Fig:3}
\end{figure*}
From Figure \ref{Fig:3}, it follows that the uncertainty relation $\lambda I_{\rho}(\sigma_1)+\frac{1}{1-\lambda} I_{\rho}(\sigma_2)$
attains the minimum value when the parameters $\lambda=\frac{1}{2}$ and $\theta=\frac{\pi}{2}$.

\section{Entanglement Detection via Uncertainty Relation based on Skew Information }
\noindent
The skew information $I_{\rho}(A)$, viewed as a quantum uncertainty of $A$ at the quantum state $\rho$, has been well studied by Lieb in \cite{Lieb}. Among various characteristic properties, the
convexity and additivity are the most important ones.

Skew information entropy is a
convex function in $\rho$, that is to say,
 if $\rho$ is a bipartite state on the quantum system $\mathcal{H}_1\otimes \mathcal{H}_2$, $\rho=\sum_{k}p_{k}\rho_{k}$ is a convex combination (i.e., $p_{k}\geq 0$, $\sum_{k}p_{k}=1$) of
 some states $\{\rho_{k}\}$, $\{M_i\}$ are some observables, then one has that
\begin{equation}\label{sep1}
\sum_i I_{\rho}(M_i)\leq \sum_{k}p_{k}\sum_{i}I_{\rho_{k}}(M_{i}).
\end{equation}

We call a state ¡°violating inequality $(\ref{sep1})$¡± iff there are no states $ \{\rho_{k}\}$ and no $\{p_k\}$ such that inequality $(\ref{sep1})$ is fulfilled.
That is different with the variance which is concave in $\rho$ on the contrary. Inequality $(\ref{sep1})$ has an obvious physical interpretation: one cannot decrease the
uncertainty of an observable by mixing several states. Moreover, in the case $\rho$ is separable, i.e., $\rho$ is a convex combination of product states, $\{\rho_{k}\}$ is
 a set of product states, violation of the inequality $(\ref{sep1})$  implies entanglement of the state; therefore, entanglement can be detected with skew information uncertainties \cite{czq}.
  Furthermore, it can be used to define the correlation limit of separable states \cite {Lieb, cai}.

The skew information entropy is fixed by the state $\rho$ and the observable $H$, Luo introduced a quantity according to skew information \cite{luo}
\begin{equation}
Q(\rho)=\sum_{i=1}^{n^2}I_{\rho}(H^{i}),
\end{equation}
where $\{H^{i}\}$ is an orthonormal basis for Hilbert space $\mathcal{L(H)}$ consisting with all observables on quantum system $\mathcal{H}$ with dimensional $n$.
Then $Q(\rho)$ is an intrinsic quantity only depending on state $\rho$, and it is independent of the choice of the orthonormal basis $\{H^{i}\}$. Also $Q(\rho)$ is
both a measure of information content of $\rho$ and a measure of quantum uncertainty.
 Then we can obtain a separability criterion depending on $Q(\rho)$.

 \begin{theorem}
Let $\rho$ be a bipartite state on the quantum system
$\mathcal{H}_1\otimes \mathcal{H}_2$, if $\rho$ is separable,
then the following inequality  holds
\begin{equation}\label{sep0}
Q(\rho)\leq \sum_{i} p_i Q(\rho_{i}),
\end{equation}
where $\sum_{i}p_i=1$, $p_i\geq 0$, $\{\rho_i\}$ are product states on $\mathcal{H}_1\otimes \mathcal{H}_2$.
\end{theorem}

%As an example, we consider the mixed state $\rho(p)=p|\varphi\rangle\langle\varphi|+\frac{1-p}{4}I$, where $|\varphi\rangle=\frac{1}{\sqrt{2}}(|00\rangle+|11\rangle)$, $0 \leq p\leq 1$. Taking the
Pauli matrices and identity matrix as the matrix basis in space of observables with dimensional 4,

The subadditivity of quantum entropy $S(\rho, M)$ describes the correlation between quantum state with its partial traces, and global measurement between the local measurement.
The entropy $S(\rho, M)$ used in this definition may be the standard Shannon entropy $S(\rho, M)=-\sum_k p_k\rm {ln}(p_k)$, or, more generally any so-called entropic function
 $S(\rho, M)=\sum_i s(p_i)$ where $s: [0, 1]\rightarrow \textbf{R}$ is a concave function, may be used.
 Let $\rho$ be a bipartite quantum state on Hilbert space $\mathcal{H}_1\otimes \mathcal{H}_2$, the partial traces of quantum state $\rho_1=\rm {Tr}_2 (\rho)$ and $\rho_2=\rm {Tr}_1 (\rho)$ are operators on
 subsystem $\mathcal{H}_1$ and $\mathcal{H}_2$, respectively.
Subadditivity of quantum entropy is stated as follows,
\begin{equation}
S(\rho, A\otimes I_2+I_1\otimes B)\leq S(\rho_1, A)+S(\rho_2, B).
\end{equation}

However, skew information is as an entropy, the subadditivity is not satisfied, that is to say,
the inequality
$
I_{\rho}(A\otimes I_2+I_1\otimes B)\leq I_{\rho_1}(A)+I_{\rho_2}(B)
$
is not hold. The state $\rho$ and that of the partial trace $\rho_1$ and $\rho_2$ have the relation \cite{Lieb}
\begin{equation}
I_{\rho}(A \otimes I_2)\geq I_{\rho_1}(A), I_{\rho}(I_1 \otimes B)\geq I_{\rho_2}(B),
\end{equation}
for arbitrary Hermitian operator $A$ in $\mathcal{H}_1$, where $I_2$ denotes the identity operator in $\mathcal{H}_2$.

Particularly, the skew information entropy has additivity in this sense,
\begin{lemma}
 Let $\rho_1$ and $\rho_2$ be two density operators of two
subsystems, and $A_1$(resp. $A_2$) be a self-adjoint operator on
subsystem $H_1$(resp. $H_2$). Let $M=A_1\otimes I_2+I_1\otimes A_2$, then
the skew information $I_{\rho}(M)$ is additive in the
sense that
if $\rho=\rho_1\otimes \rho_2$, then
$I_{\rho}(M) =I_{\rho_1}(A_1)+I_{\rho_2}(A_2)$, where $I_1$ and
$I_2$ are the density matrices for the first and second systems,
respectively.
\end{lemma}
 \textbf{Proof:} Suppose $\rho=\rho_1\otimes \rho_2$, then
 \begin{equation}
\begin{split}
\sqrt{\rho}M\sqrt{\rho}M&=\sqrt{\rho_1}A_1\sqrt{\rho_1}A_1\otimes \rho_2+\sqrt{\rho_1}A_1\sqrt{\rho_1}\\
&\otimes\rho_2A_2+\rho_1A_1\otimes \sqrt{\rho_2}A_2\sqrt{\rho_2}\\
&+\rho_1\otimes \sqrt{\rho_2}A_2\sqrt{\rho_2}A_2,
\end{split}
\end{equation}
 and
  \begin{equation}\label{a1}
\begin{split}
 \rm {Tr}(\sqrt{\rho}&M\sqrt{\rho}M)=\rm {Tr}(\sqrt{\rho_1}A_1\sqrt{\rho_1}A_1)\\
 &+\rm {Tr}(\sqrt{\rho_2}A_2\sqrt{\rho_2}A_2)
 +2\rm {Tr}(\rho_1A_1)Tr(\rho_2A_2),
 \end{split}
\end{equation}
also,
\begin{equation}\label{a2}
\begin{split}
\rm {Tr}(\rho
M^2)&=\rm {Tr}(\rho_1A_1^2)+\rm {Tr}(\rho_2A_2^2)\\
&+2\rm {Tr}(\rho_1A_1)\rm {Tr}(\rho_2A_2).
\end{split}
\end{equation}
It follows from \eqref{a1}--\eqref{a2} that
$I_{\rho}(M)=\rm {Tr}(\rho
M^2)-\rm {Tr}(\sqrt{\rho}M\sqrt{\rho}M)=I_{\rho_1}(A_1)+I_{\rho_2}(A_2)$.

Now we consider the following scenario: Alice and Bob perform local measurements $A_i$ and $B_i$, $i=1,2,...,k$ on
 an unknown quantum state $\rho$ respectively. Their job is to judge if $\rho$ is entangled or not. And the ``sum observables'' $M_i$ are given by
 \begin{equation}
 M_i=A_i\otimes I_2+I_1\otimes B_i.
 \end{equation}
A separability criterion based on the local uncertainty relations is obtained.

\begin{theorem} If quantum state $\rho$ is separable, then the following inequality holds
\begin{equation}\label{sep2}
\sum_i I_{\rho}(M_i)\geq c_{A}+c_{B},
\end{equation}
where $c(A)$ and $c(B)$ are the optimal uncertainty constants for
the observables $\{A_i\}$ and $\{B_i\}$, i.e, $\sum_i
I_{\rho_{A}}(A_i)\geq c_{A}$, $\sum_i I_{\rho_{B}}(B_i)\geq c_{B}$.
\end{theorem}
\textbf{Proof:}
Let $\rho=\rho_{A}\otimes \rho_{B}$ be a product state, and  $\sum_i I_{\rho_{A}}(A_i)\geq c_{A}$,
$\sum_i I_{\rho_{B}}(B_i)\geq c_{B}$, since the skew information is additive, the inequality $(\ref{sep2})$ holds.
Because of the convexity of skew information, this inequality also holds for all convex combinations of product states \cite{Lieb}, i.e., for all separable states.

Inequality $(\ref{sep2})$ manifests the correlation between the sum uncertainty and the local uncertainty for separable states.
Any violation of the limit of the uncertainty therefore proves that
the quantum state cannot be separated into a mixture of product
states. The violation of any local uncertainty relation of
the form $(\ref{sep2})$ is therefore a sufficient condition for the existence
of entanglement.

Furthermore, relation $(\ref{sep2})$ is a
spin-squeezing \cite{squ} criterion for the angular momentum measurements. As such, it requires the same experimental data as other spin-squeezing criteria, see Refs.
\cite{spin1, spin2}, namely, only a measurement of first and second moments of the total angular momentum In
contrast to entanglement criteria based on tomography,
these are advantageous in typical experimental implementation.

\section{Conclusions}
\noindent
Uncertainty relations are one of the central properties in quantum theory and quantum information. We have investigated
the uncertainty relation based on the Wigner-Yanase skew information for multiple noncommutative observables. The
corresponding lower bounds derived in this paper are shown to be tighter than the
previous ones, and thus capture better the incompatibility of the observables. The results are expected to shed new lights on investigating  quantum task, as uncertainty
relations are closely related to many quantum information processing like entanglement detection, security analysis of quantum key distribution in quantum cryptography and non-locality quantum tasks.

\bigskip
\noindent{\bf Acknowledgments}. This work is supported by the
National Natural Science Foundation of China (grant Nos. 11861031
and 11531004) and Simons Foundation grant No. 523868.

\end{multicols}

\end{document}